\documentclass{elsart}
\usepackage{graphicx,cite}
\usepackage{amssymb}
\usepackage{ulem}
\usepackage[dvipdfmx,bookmarks=false]{hyperref}%

\newcommand{\affuni}[2]{Dipartimento di Fisica dell'Universit\`a
  #1, #2, Italy.}
\newcommand{\affinfn}[2]{INFN Sezione #1, #2, Italy.}

\def\ff{$\phi$--factory}  \def\DAF{DA\char8NE}

\newcommand{\bfi}{\begin{figure}}
\newcommand{\efi}{\end{figure}}
\newcommand{\ba}{\begin{eqnarray}}
\newcommand{\ea}{\end{eqnarray}}
\newcommand{\be}{\begin{equation}}
\newcommand{\ee}{\end{equation}}
\newcommand{\bnona}{\begin{eqnarray*}}
\newcommand{\enona}{\end{eqnarray*}}

\newcommand{\dd}{{\rm d}}
\newcommand{\MeV}{\mathrm{\ MeV}}
\newcommand{\GeV}{\mathrm{\ GeV}}

\def\ifm#1{\relax\ifmmode#1\else$#1$\fi}  \def\plm{\ifm{\pm}}
\def\km{\kern-1.5mm}  \def\kak{\km&\km}  \def\kma{\kern-2.5mm}\def\kms{\kern-.75mm}

\def\bbm[#1]{\mbox{\boldmath $#1$}}
\def\up#1{$^{#1}$}     \def\deg{\ifm{^\circ}}
\def\ie{{\it\kern-2pt i.\kern-.5pt e.\kern3pt}}  \def\gam{\ifm{\gamma}}  \def\mmu{\ifm{\mu\mu}}
\def\eg{{\it\kern-2pt e.\kern-.5pt g.\kern3pt}}  \def\dif{\hbox{d}}   \def\sig{\ifm{\sigma}}
\def\epm{\ifm{e^+e^-}}  \def\to{\ifm{\rightarrow}}  \def\pic{\ifm{\pi^+\pi^-}}  \def\po{\ifm{\pi^0}}
\def\x{\ifm{\times}}  \def\ab{\ifm{\sim}}   \def\pt#1;#2;{\ifm{#1\x10^{#2}}}

\begin{document}

\begin{frontmatter}

\title{\boldmath Precision measurement of $\sigma(e^+e^-\rightarrow\pi^+\pi^-\gamma)/\sigma(e^+e^-\rightarrow \mu^+\mu^-\gamma)$ and determination of the $\pi^+\pi^-$ contribution to the muon anomaly with the KLOE detector}

\collab{The KLOE and KLOE-2 Collaborations}
\author[Frascati]{D.~Babusci},
\author[Roma2,INFNRoma2]{D.~Badoni},
\author[Cracow]{I.~Balwierz-Pytko},
\author[Frascati]{G.~Bencivenni},
\author[Roma1,INFNRoma1]{C.~Bini},
\author[Frascati]{C.~Bloise},
\author[Frascati]{F.~Bossi},
\author[INFNRoma3]{P.~Branchini},
\author[Roma3,INFNRoma3]{A.~Budano},
\author[Uppsala]{L.~Caldeira~Balkest\aa hl},
\author[Frascati]{G.~Capon},
\author[Roma3,INFNRoma3]{F.~Ceradini},
\author[Frascati]{P.~Ciambrone},
\author[Messina,INFNMessina]{F.~Curciarello},
\author[Frascati]{E.~Czerwi\'nski},
\author[Frascati]{E.~Dan\'e},
\author[Messina,INFNMessina]{V.~De~Leo},
\author[Frascati]{E.~De~Lucia},
\author[INFNBari]{G.~De~Robertis},
\author[Roma1,INFNRoma1]{A.~De~Santis},
\author[Frascati]{P.~De~Simone},
\author[Roma1,INFNRoma1]{A.~Di~Domenico},
\author[Napoli,INFNNapoli]{C.~Di~Donato},
\author[Frascati]{D.~Domenici},
\author[Bari,INFNBari]{O.~Erriquez},
\author[Bari,INFNBari]{G.~Fanizzi},
\author[Frascati]{G.~Felici},
\author[Roma1,INFNRoma1]{S.~Fiore},
\author[Roma1,INFNRoma1]{P.~Franzini},
\author[Roma1,INFNRoma1]{P.~Gauzzi},
\author[Messina,INFNMessina]{G.~Giardina},
\author[Frascati]{S.~Giovannella},
\author[Roma2,INFNRoma2]{F.~Gonnella},
\author[INFNRoma3]{E.~Graziani},
\author[Frascati]{F.~Happacher},
\author[Uppsala]{L.~Heijkenskj\"old},
\author[Uppsala]{B.~H\"oistad},
\author[Frascati]{L.~Iafolla},
\author[Energetica,Frascati]{E.~Iarocci},
\author[Uppsala]{M.~Jacewicz},
\author[Uppsala]{T.~Johansson},
\author[Karlsruhe]{W.~Kluge},
\author[Uppsala]{A.~Kupsc},
\author[Frascati,StonyBrook]{J.~Lee-Franzini},
\author[INFNBari]{F.~Loddo},
\author[Frascati,BINP]{P.~Lukin},
\author[Messina,INFNMessina,centro]{G.~Mandaglio},
\author[Moscow]{M.~Martemianov},
\author[Frascati,Marconi]{M.~Martini},
\author[Roma2,INFNRoma2]{M.~Mascolo},
\author[Roma2,INFNRoma2]{R.~Messi},
\author[Frascati]{S.~Miscetti},
\author[Frascati]{G.~Morello},
\author[INFNRoma2]{D.~Moricciani},
\author[Cracow]{P.~Moskal},
\author[Frascati,KVI]{S.~M\"uller},
\author[INFNRoma3,LIP]{F.~Nguyen\corauthref{cor}},
\ead{federico.nguyen@cern.ch}
\corauth[cor]{Corresponding authors.}
\author[INFNRoma3]{A.~Passeri},
\author[Energetica,Frascati]{V.~Patera},
\author[Roma3,INFNRoma3]{I.~Prado~Longhi},
\author[INFNBari]{A.~Ranieri},
\author[Uppsala]{C.~F.~Redmer},
\author[Frascati]{P.~Santangelo},
\author[Frascati]{I.~Sarra},
\author[Calabria,INFNCalabria]{M.~Schioppa},
\author[Frascati]{B.~Sciascia},
\author[Cracow]{M.~Silarski},
\author[Roma3,INFNRoma3]{C.~Taccini},
\author[INFNRoma3]{L.~Tortora},
\author[Frascati]{G. Venanzoni\corauthref{cor}},
\ead{graziano.venanzoni@lnf.infn.it}
\author[Frascati,CERN]{R.~Versaci},
\author[Warsaw]{W.~Wi\'slicki},
\author[Uppsala]{M.~Wolke},
\author[Cracow]{J.~Zdebik}.

\address[Bari]{\affuni{di Bari}{Bari}}
\address[INFNBari]{\affinfn{Bari}{Bari}}
\address[Calabria]{\affuni{della Calabria}{Cosenza}}
\address[INFNCalabria]{INFN Gruppo collegato di Cosenza, Cosenza, Italy.}
\address[centro]{Centro Siciliano di Fisica Nucleare e Struttura della Materia, Catania, Italy.}
\address[INFNMessina]{\affinfn{Catania}{Catania}}
\address[Cracow]{Institute of Physics, Jagiellonian University, Cracow, Poland.}
\address[Frascati]{Laboratori Nazionali di Frascati dell'INFN, Frascati, Italy.}
\address[Karlsruhe]{Institut f\"ur Experimentelle Kernphysik, Universit\"at Karlsruhe, Germany.}
\address[Messina]{Dipartimento di Fisica e Scienze della Terra dell'Universit\`a di Messina, Messina, Italy.}
\address[Moscow]{Institute for Theoretical and Experimental Physics (ITEP), Moscow, Russia.}
\address[Napoli]{\affuni{''Federico II''}{Napoli}}
\address[INFNNapoli]{\affinfn{Napoli}{Napoli}}
\address[Energetica]{Dipartimento di Scienze di Base ed Applicate per l'Ingegneria dell'Universit\`a
``Sapienza'', Roma, Italy.}
\address[Marconi]{Dipartimento di Scienze e Tecnologie applicate, Universit\`a "Guglielmo Marconi", Roma, Italy.}
\address[Roma1]{\affuni{''Sapienza''}{Roma}}
\address[INFNRoma1]{\affinfn{Roma}{Roma}}
\address[Roma2]{\affuni{``Tor Vergata''}{Roma}}
\address[INFNRoma2]{\affinfn{Roma Tor Vergata}{Roma}}
\address[Roma3]{\affuni{``Roma Tre''}{Roma}}
\address[INFNRoma3]{\affinfn{Roma Tre}{Roma}}
\address[StonyBrook]{Physics Department, State University of New
York at Stony Brook, USA.}
\address[Uppsala]{Department of Physics and Astronomy, Uppsala University, Uppsala, Sweden.}
\address[Warsaw]{National Centre for Nuclear Research, Warsaw, Poland.}
\address[CERN]{Present Address: CERN, CH-1211 Geneva 23, Switzerland.}
\address[KVI]{Present Address: KVI, 9747 AA Groningen, The Netherlands.}
\address[LIP]{Present Address: Laborat\'orio de Instrumenta\c{c}\~{a}o e F\'isica Experimental de Part\'iculas,
Lisbon, Portugal.}
\address[BINP]{Present Address: Budker Institute of Nuclear Physics, 630090 Novosibirsk, Russia.}
\vspace{8mm}
\begin{abstract}
We have measured the ratio $\sigma(e^+e^-\rightarrow\pi^+\pi^-\gamma)/\sigma(e^+e^-\rightarrow \mu^+\mu^-\gamma)$, with the KLOE detector at \DAF\ for a total integrated luminosity of \ab240 pb$^{-1}$.
From this ratio we obtain the cross section $\sigma(e^+e^-\rightarrow\pi^+\pi^-)$.
From the cross section we determine the pion form factor 
$|F_\pi|^2$ 
and the two-pion contribution to the muon anomaly $a_\mu$ for $0.592<M_{\pi\pi}<0.975$ GeV, {\bf $\Delta^{\pi\pi} a_\mu$= $({\rm 385.1\pm1.1_{stat}\pm2.7_{sys+theo})}\times10^{-10}$}. This result confirms the current discrepancy between the Standard Model calculation and the experimental measurement of the muon anomaly.
 \end{abstract}



\vskip -1.5cm 
\end{frontmatter}

\section{Introduction}
\label{sec:1}
Measurements of the muon magnetic anomaly $a_\mu=(g_\mu-2)/2$ performed at the Brookhaven Laboratory
have reached an accuracy of 0.54 ppm: $a_\mu=\pt(11\,659\,208.9\pm6.3);-10;$ \cite{Bennett:2006fi,Beringer:1900zz}. The quoted value differs from Standard Model estimates by 3.2-3.6 standard deviations \cite{Miller:2007kk,Jegerlehner:2009ry,Davier:2010nc,Hagiwara:2011af}\footnote{A recent evaluation \cite{Benayoun:2012wc} finds a difference between 4.7 and 4.9 standard deviations.}.
The difference between measurement and calculations is of great interest since it could be a signal of New Physics. The authors of Ref. \citen{Czarnecki:2001pv} have proposed an interpretation in terms of Supersymmetry, which can be probed at the Large Hadron Collider. Another proposal suggests the existence of a light vector boson in the Dark Matter sector, coupled with ordinary fermions through photon exchange, which is not excluded by present low energy tests of the Standard Model \cite{ArkaniHamed:2008qn,Pospelov:2008zw}. A new round of measurements of $a_\mu$ is expected
at Fermilab \cite{Carey:2009zz} and J-PARC \cite{Imazato:2004fy}, with the aim of considerably reducing the experimental error. To fully exploit the significance of improved measurements of $a_\mu$ it is important to confirm the present estimate of the hadronic corrections (see below) and possibly to decrease 
the corresponding error.

The main source of uncertainty in the Standard Model estimates of $a_\mu$ \cite{Miller:2007kk,Jegerlehner:2009ry}
is due to hadronic loop contributions which are not calculable in perturbative QCD. To lowest order, the hadronic contribution $\Delta^{\rm h,\,lo}a_\mu$, can be obtained from a dispersion integral \cite{mich,deraf} over the ``bare'' cross section \sig\up0(\epm\to\ hadrons(\gam)). \sig\up0 is obtained from the physical cross section, inclusive of final state radiation, removing vacuum polarization, VP, effects and contributions due to additional photon emission in the initial state.
The leading-order hadronic contribution is \ab\pt690;-10;, the precise value depending on the authors' different averaging procedures, as discussed in Refs. \citen{Miller:2007kk,Jegerlehner:2009ry,Davier:2010nc,Hagiwara:2011af}.
The \epm\to\pic(\gam)\ process contributes approximately $75\%$ of the $\Delta^{\rm h,\,lo}a_\mu$ value and accounts for about $40\%$ of its uncertainty.

In the following, we discuss the measurement of the cross sections as a function of the $\mu^+\mu^-$ and  $\pi^+\pi^-$ invariant masses $M_{\mu\mu}$ and $M_{\pi\pi}$:
$${\dif\sigma(\epm\to\mu^+\mu^-\gam)\over\dif s_\mu}\qquad \hbox{and}\qquad  {\dif\sigma(\epm\to\pic\gam)\over\dif s_\pi}$$
with $s_\mu=M^2_{\mu\mu},\ s_\pi=M^2_{\pi\pi}$, to be used for the determination of $\sigma^0(\epm\to\pic)$. From the latter we obtain the two-pion contribution to the anomaly, $\Delta^{\rm h,\,lo}a_\mu$ and the pion form factor $|F_\pi|^2$ for comparison to other results.
\section{Measurement of $\sigma(\pi^+\pi^-)$ at \DAF}
\label{sec:2}
The KLOE detector operates at \DAF, the Frascati \ff, an $e^+e^-$ collider running at fixed energy, $W=\sqrt s\sim1020\MeV$, the $\phi$ meson mass. Initial state radiation (ISR) provides a means to produce \pic\ pairs of variable $s_\pi$.
Counting \pic\gam\ events leads to a measurement of $\dif\sigma(\epm\to\pic\gam)/\dif s_\pi$ if the integrated luminosity is known, from which  \sig(\epm\to\pic) can be extracted.
We have published three measurements \cite{Ambrosino:2010bv,Ambrosino:2008aa,Aloisio:2004bu}
of \sig(\epm\to\pic) for $0.1 <M_{\pi\pi}^2<0.95\GeV^2$, with results consistent within errors and a combined fractional uncertainty of about $1\%$. The luminosity was obtained by counting Bhabha scattering events and using the QED value of the corresponding cross section.
\bfi[htb]
\centering
  \includegraphics[width=.8\textwidth]{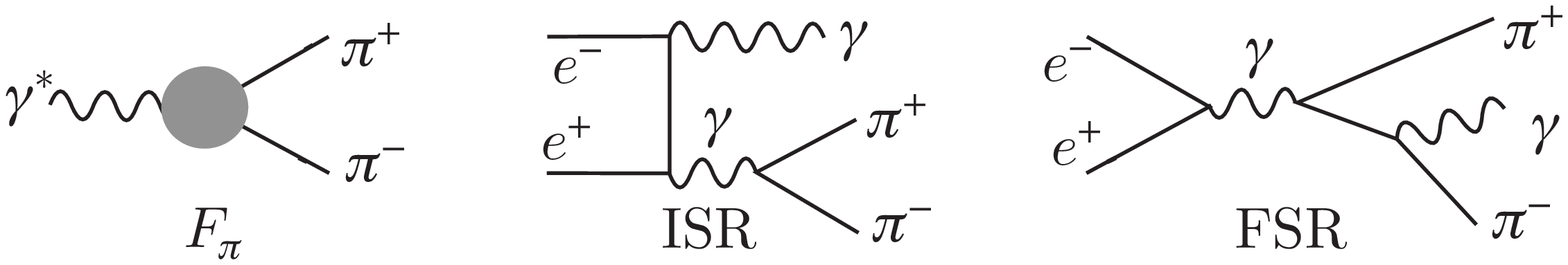}
  \caption{Simplified amplitudes for $\gamma^*\to\pic$, \ \epm\to\pic\gam\ (ISR) and \epm\to\pic\gam\ (FSR).}
\label{ifsr}
\efi
To lowest order, the pion form factor is defined by:
\be
\langle\pic|J^{\rm em}_\mu(\pi)|0\rangle=(p_{\pi^+}-p_{\pi^-})_\mu\x F_\pi\left(s_\pi=(p_{\pi^+}+p_{\pi^-})^2\right).
\ee
where $p_{\pi^+}$, $p_{\pi^-}$ are the momenta of $\pi^+$ and $\pi^-$.
The  differential cross section  for $e^+e^-\to\pi^+\pi^-\gamma$ 
due to the ISR amplitude of Fig. \ref{ifsr}
is related to the dipion cross section
 $\sigma_{\pi\pi}\equiv \sigma(e^+e^-\to\pi^+\pi^-\gamma)$ \cite{Binner:1999bt}:

\begin{equation}
s \left. \frac{\dd\sigma(\pi^+\pi^-\gamma)} {\dd s_\pi}\right|_{{\rm ISR}}=
\sigma_{\pi\pi}(s_\pi)~ H(s_\pi,s),
\label{eq:1}
\end{equation}
where the radiator function $H$ is computed from QED with 
complete NLO corrections\cite{Rodrigo:2001kf,Czyz:2002np,Czyz:2003ue,Czyz:2004rj,Actis:2010gg} 
and depends on the  $e^+e^-$ center-of-mass energy squared $s$. $\sigma_{\pi\pi}$ obtained from Eq. \ref{eq:1} requires accounting for final state radiation
(FSR in Fig. \ref{ifsr}). 
In the following we only use events where the photon is emitted at small angles, as discussed in detail 
in Refs. \citen{Ambrosino:2008aa,Aloisio:2004bu}.
The cross section for \epm\to\pic\gam\ is proportional to the two-photon \epm\ annihilation cross section, which diverges, at lowest order, for the photon angle going to zero. This is not the case for the FSR contribution. Our choice results in a large enhancement of the ISR with respect to the FSR contribution.

Equation \ref{eq:1} is also valid for $\epm\to\mu^+\mu^-\gam$ and $\epm\to\mu^+\mu^-$ with the same radiator function $H$. We can therefore determine $\sigma_{\pi\pi}$ from the ratio of the
$\pi^+\pi^-\gamma$ and $\mu^+\mu^-\gam$ differential cross sections for the same value of the dipion and dimuon invariant mass (see also Refs. \citen{Aubert:2009ad,Lees:2012cj}). For ISR events we have:
\be\label{eq:3}\sigma^0(\pic,\ s')={\dif\sigma(\pic\gam,{\rm\ ISR})/\dif s'\over\dif\sigma(\mu^+\mu^-\gam,{\rm\ ISR})/\dif s'}\:\x\,\sigma^0(\epm\to\mu^+\mu^-,\ s').\ee
where $s'= s_{\pi} = s_{\mu}$.

Final state photon 
emission for both the $\pi^+\pi^-\gamma$ and $\mu^+\mu^-\gamma$ channels slightly modifies Eq. \ref{eq:3}. These corrections are included in our analysis \cite{knratio}.

From the bare cross section $\sigma^0_{\pi\pi}$ we obtain the pion form factor:
\be\label{eq:4}
|F_\pi(s')|^2 = \frac{3}{\pi}\frac{s'}{\alpha^2\beta^3_\pi}
\,\,\sigma^0_{\pi\pi(\gamma)}(s')(1+\delta_{VP})(1-\eta_{\pi}(s'))\ee
where $\delta_{VP}$ is the VP correction\cite{fj} and
$\eta_\pi$ accounts for FSR radiation assuming point-like pions\cite{Schwinger:1989ix}.

The advantages of the ratio method are:
\begin{enumerate}
\item the $H$ function does not appear in Eq. \ref{eq:3}. Therefore the measurement of $\sigma_{\pi\pi}$ is not affected by the related systematic uncertainty of $0.5\%$ \cite{Rodrigo:2001kf} ;
\item using the same data sample for the $\pi^+\pi^-\gamma$ and $\mu^+\mu^-\gamma$
events, there is no need for luminosity measurements;
\item vacuum polarization corrections and most other radiative corrections cancel in the ratio;
\item using the same fiducial volume, acceptance corrections
to the $\pi^+\pi^-\gamma$ and $\mu^+\mu^-\gamma$ spectra almost cancel resulting in a small
systematic uncertainty.
\end{enumerate}
In the following we describe the measurement of $\dd\sigma_{\mu\mu\gamma}/\dd s_\mu$ using the same
data as those to measure $\dd\sigma_{\pi\pi\gamma}/\dd s_\pi$ \cite{Ambrosino:2008aa}. 

\section{Measurement of the \bbm[e^+e^-\to\mu^+\mu^-\gamma] cross section}
\label{sec:3}
The data sample corresponds to an integrated luminosity of 239.2 pb$^{-1}$
collected in 2002, with low machine background and stable \DAF\
conditions. We also recorded events without offline filters with a downscaled trigger, providing control samples for evaluating efficiencies.
\subsection{The KLOE Detector}
\label{subsec:3.1}
\begin{figure}[htbp]\centering
  \includegraphics[width=.45\textwidth]{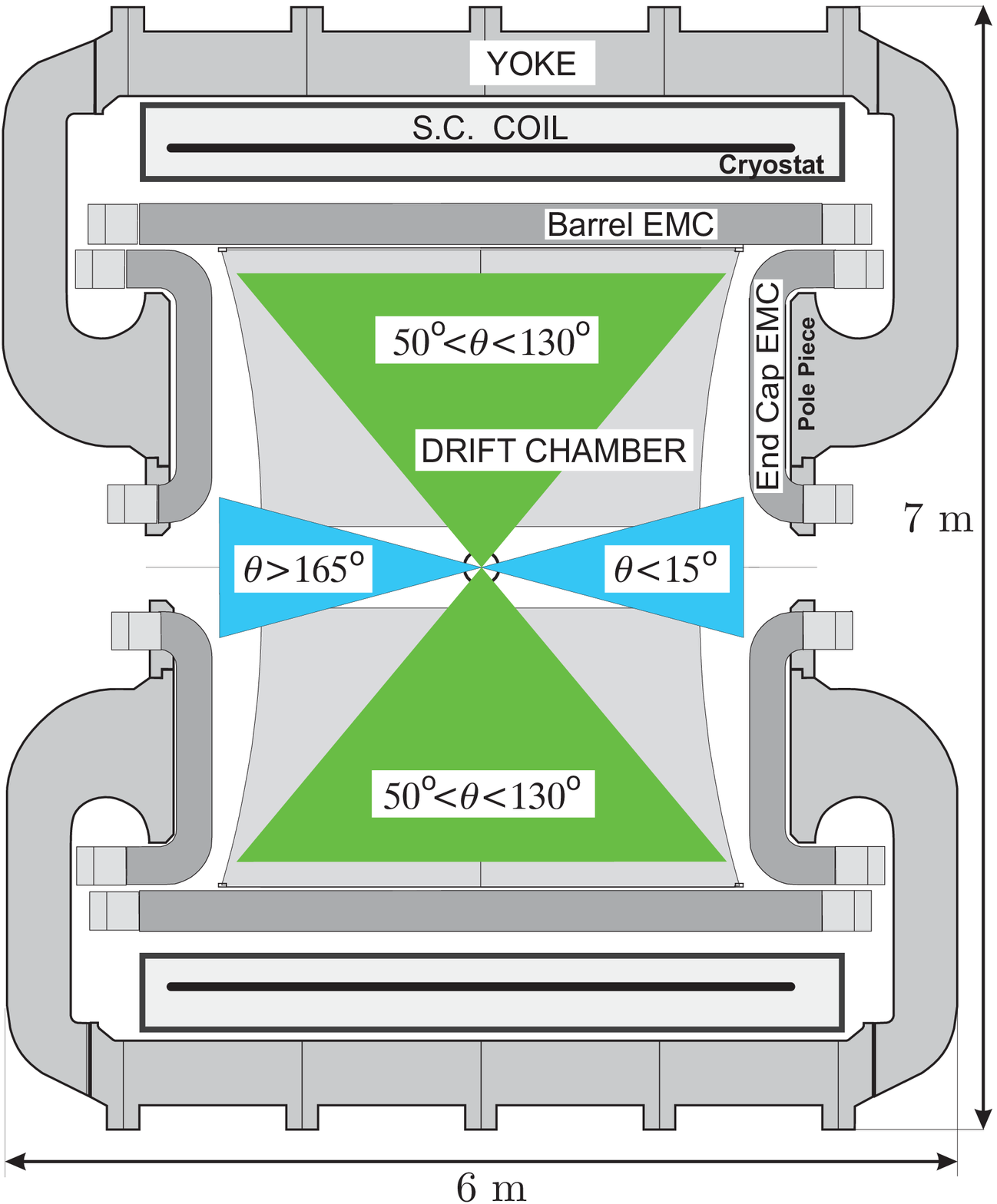}
  \caption{Vertical cross section of the KLOE detector,
    showing the small and large angle regions where respectively photons and muons are accepted}
\label{fig:1}\end{figure}

The KLOE detector consists of a cylindrical drift chamber (DC) \cite{Adinolfi:2002uk}
and a lead--scintillating fibers electromagnetic calorimeter (EMC) \cite{Adinolfi:2002zx}.
The DC has a momentum resolution of $\sigma_{p_\bot}/p_\bot\sim 0.4\%$ for tracks with polar angle $\theta\!>\!45^\circ$.
Track points are measured in the DC with a resolution in $r$-$\phi$ of about $0.15$ mm and about $2$ mm in $z$.
The EMC has an energy resolution of $\sigma_E/E\sim 5.7\%/\sqrt{\rm E\ (GeV)}$ and an excellent time
resolution of $\sigma_t\sim 57\ {\rm ps}/\sqrt{\rm E\ (GeV)}\oplus 100\ {\rm ps}$. A cross section of the detector in the $y,\:z$ plane is shown in Fig. \ref{fig:1}.
A superconducting coil provides an axial magnetic field of 0.52 T along the bisector of the colliding beam directions. The bisector is taken as the $z$ axis of our coordinate system.
The $x$ axis is horizontal, pointing to the center of the collider rings and the $y$ axis is vertical, directed upwards.
The trigger~\cite{Adinolfi:2002hs} uses both EMC and DC information.
Events used in this analysis are triggered by two energy deposits larger than 50 MeV
in two sectors of the barrel calorimeter.

\subsection{Identification of $e^+e^-\to\mu^+\mu^-\gamma$ events}\label{subsec:3.2}
The signature for  $e^+e^-\to\mu^+\mu^-\gamma$ events with the photon emitted at small angle is just two tracks of
opposite curvature, the photon being lost in the beam pipe. Four types of events contribute to the above signature:
1: $e^+e^-\to\mu^+\mu^-\gamma$, 2: \epm\to\pic\gam\ , 3: \epm\to\epm\gam, and 4: \epm\to\pic\po . The four reactions can be distinguished kinematically. 
From 
 the overdetermined system of kinematical
constraints of the reaction $\epm\to x^+x^-\gam$,
 we can compute the common mass $(m_x)$ of particles $x^+$ and $x^-$. 
The four processes give $m_x=m_\mu,\ m_\pi,\ m_e$ and $>m_{\pi}$. Additional separation between electrons and pions or muons 
is obtained from a particle identification (PID) estimator for each track, $L_\pm$, 
which uses time-of-flight information and the value and shape of the energy deposit of each charged particle
 in the calorimeter \cite{Ambrosino:2008aa}. Figure \ref{fig:1} shows the fiducial volume we use for muons and unobserved photons which is identical to that used in Ref. \citen{Ambrosino:2008aa} for \epm\to\pic\gam.

\begin{figure}[htb]\centering \includegraphics[width=.55\textwidth]{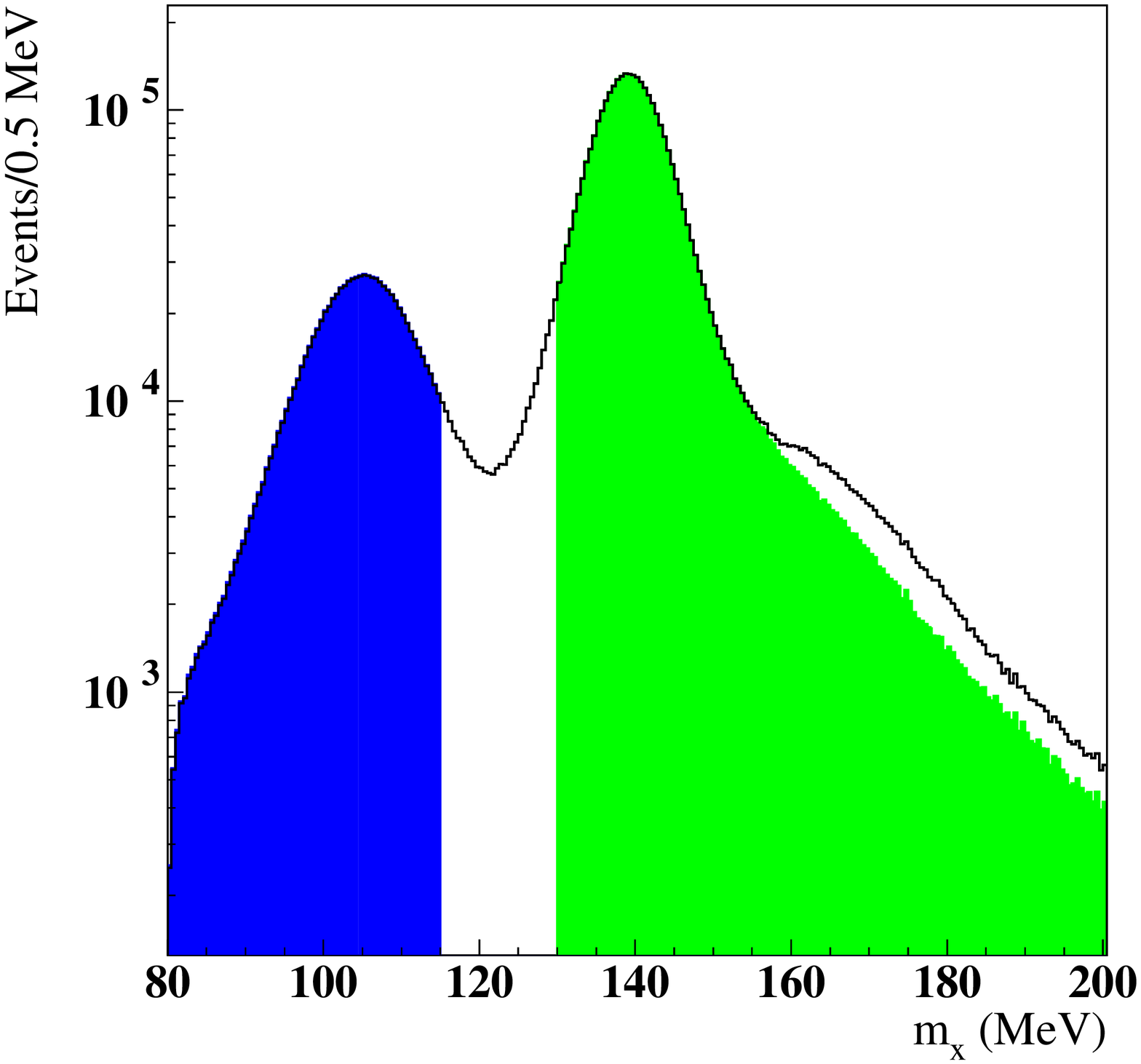}\caption{Data \mmu\gam\  and $\pi\pi\gamma$  regions in the $m_x$ spectrum. 
The \mmu\gam\  and  $\pi\pi\gamma$ accepted regions are shown in
blue and green.
A residual contamination of \pic\po\ events is visible at high $m_x$ values.}
\label{fig:2}\end{figure}

We list below the requirements for \mmu\gam\ event acceptance.
\begin{enumerate}
\item Events must have 
at least 
 two tracks of opposite sign, with origin at the interaction point and polar angle satisfying $50\deg<\theta<130\deg$. The reconstructed momenta must satisfy $p_\bot>160$ MeV or $|p_z|>90$ MeV, to ensure good reconstruction and efficiency.
\item The polar angle  $\theta_{\mu\mu}$ of the dimuon system obtained from the momentum of the two tracks (${\bf p}_{\mu\mu}={\bf p}_++ {\bf p}_-$) must satisfy $|\cos\theta_{\mu\mu}|>\!\cos(15\deg)$.
\item Events with both tracks having $L_\pm<$~0 are identified as $ee\gam$ events and rejected.
The loss due to this cut is less than $0.05\%$, as evaluated with \mmu\ samples, obtained from both data and Monte Carlo.
\item The computed mass for the two observed particles must satisfy $80<m_x<115\MeV$.
\end{enumerate}
About $8.9\times10^5$ \mmu\gam\ events pass these criteria, while $34.9\times10^5$ events are selected as $\pi\pi\gamma$\cite{Ambrosino:2008aa}.
Figure \ref{fig:2} shows the $m_x$ distribution together with the accepted regions for \mmu\gam\ 
and $\pi\pi\gamma$ events.

\subsection{Background estimates}
\label{subsec:3.3}
Residual $\pi^+\pi^-\gamma$, $\pi^+\pi^-\pi^0$ and $e^+e^-\gamma$ backgrounds
are evaluated by fitting the observed $m_x$ spectrum 
 with a superposition of Monte Carlo simulation (MC) distributions describing signal and
$\pi^+\pi^-\gamma$, $\pi^+\pi^-\pi^0$ backgrounds, and a 
distribution obtained from data for
the $e^+e^-\gamma$ background.
The normalized contributions from
signal and backgrounds are free parameters of the fit, performed for 
30 intervals in $M_{\mu\mu}^2$ of $0.02\GeV^2$ width for $0.35<M_{\mu\mu}^2<0.95\GeV^2$.

\begin{figure}[htb]\centering
  \includegraphics[width=.5\textwidth,height=.5\textwidth]{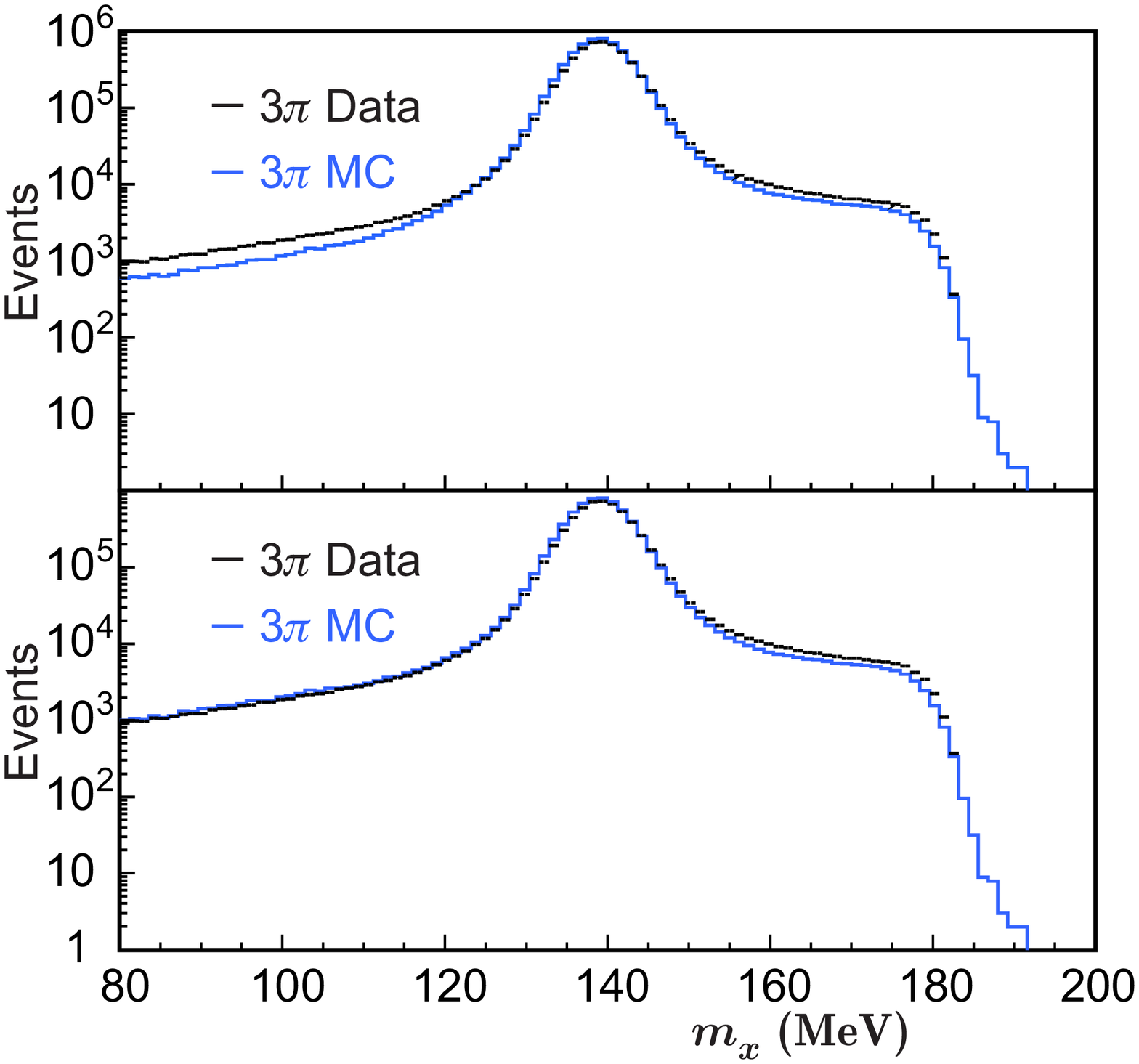}
  \caption{Data and MC $m_x$ distributions for the $\pi^+\pi^-\pi^0$ control sample, before (upper)
  and after (lower) the resolution correction.}\label{fig:3}\end{figure}

In the $\rho$ mass region, the fractional $\pi^+\pi^-\gamma$ yield in the \mmu\gam\ acceptance region is about $15\%$ of the sample.
To improve the MC description of the low energy $m_x$ tail of $\pi^+\pi^-\gamma$
events in the muon peak, Fig. \ref{fig:2}, we apply a data/MC resolution correction, function of
$s_\mu$.
This correction is evaluated from a high purity sample of $\phi\to\pi^+\pi^-\pi^0$ events,
with the same track requirements used for \mmu\gam\ events, requiring
in addition two photons with 
 an invariant mass compatible with the $\pi^0$ mass, both for data and MC~\cite{knratio}.
Figure \ref{fig:3} shows the comparison between data and MC before and after the resolution correction.

 The $\pi^+\pi^-\gamma$, $\pi^+\pi^-\pi^0$ and $e^+e^-\gamma$  background fractions as obtained by the fitting procedure are
given in Table~\ref{tab:x} at three $s_{\mu}$  values. The estimated $\pi^+\pi^-\pi^0$ fraction obtained from the fit procedure is consistent with the one expected from the theoretical 
cross section.
Contributions from $e^+ e^- \to e^+e^-\mu^+\mu^-$ and
$e^+ e^- \to e^+e^-\pi^+\pi^-$ processes are evaluated using the
\texttt{Nextcalibur}~\cite{Berends:2000fj}
 and \texttt{Ekhara}~\cite{Czyz:2006dm} MC generators.
After analysis cuts, the $e^+ e^- \to e^+e^-\pi^+\pi^-$ process is found to be negligible, while the
$e^+ e^- \to e^+e^-\mu^+\mu^-$ background contribution is between $0.6\%$ and $0.1\%$,
in the low $M_{\mu\mu}^2$ region and is subtracted from the data spectrum~\cite{knratio}.

\begin{table}[htbp]
\begin{center}
\renewcommand{\arraystretch}{1.0}
\begin{tabular}{||l|c|c|c|}
\hline\hline
Background process & 0.405 &0.605 & 0.905 \\
\hline\hline
$\pi^+\pi^-\gamma$  & $3.44\pm0.11$& $11.61\pm 0.14$ & $1.60\pm 0.03$ \\
 $\pi^+\pi^-\pi^0$  & $1.28\pm0.16$& $0.19\pm0.05$ & $<0.01$ \\
 $e^+e^-\gamma$  & $1.52\pm0.04$& $2.05\pm0.04$ & $1.73\pm 0.02$ \\
\hline\hline
\end{tabular}
\caption{List of main background fractions (in \%) for three different $s_{\mu}$ values (in GeV$^2$).}
\label{tab:x}
\end{center}
\end{table}

Systematic errors in the background subtraction include: ({\it i})
\ Errors on the parameters from the 
fit procedure: 
these decrease monotonically from $0.7\%$ to $0.1\%$  with respect to $s_\mu$; 
({\it ii}) The uncertainty on the data/MC resolution corrections: about $1\%$ in the $\rho$ mass region,
smaller at higher $s_\mu$, negligible at lower $s_\mu$ values;
({\it iii})  The uncertainty on the $e^+ e^- \to e^+e^-\mu^+\mu^-$ process: about $0.4\%$
at low $s_\mu$ values, rapidly falling to $0.1\%$ for $s_\mu>0.5\GeV^2$.
The correctness of the background estimate has been checked by two independent methods.
\begin{enumerate}
\item We perform a kinematic fit of the two-track events assuming it is a $\mu\mu\gamma$ state. The $\chi^2$
value obtained  is used as a discriminant variable, instead of  $m_x$, in the fitting procedure described above.
\item We improve the $\pi$-$\mu$ separation by use of $m_{x}$, applying a quality cut on the helix fit for both
tracks. This cut reduces the dipion background in the dimuon signal region by more than a factor of two.
\end{enumerate}
The background fractions obtained for both cases are in good agreement with the standard procedure\cite{knratio}.
\subsection{Efficiencies, acceptance and systematic errors}
\label{subsec:3.4}
The MC generator \texttt{Phokhara}, including next-to-leading-order ISR as well as FSR corrections~\cite{Czyz:2004rj}
has been inserted in the standard KLOE MC \texttt{Geanfi}~\cite{Ambrosino:2004qx}.
We compared MC efficiencies with efficiencies obtained from data control samples, and
studied two major effects: the EMC response to muons clusters and the muon DC tracking efficiency.

{\bf EMC response.}
From a subsample of $\mu\mu\gamma$ events with both tracks fitted,
the 
efficiency to find at least a cluster with $L>$0 is found to be equal to one 
within 10$^{-4}$.
The trigger efficiency is obtained from  a sample of $\mu\mu\gamma$ events  
where a single muon satisfies the trigger requirements. Then, the trigger response for the other
muon is parametrized as a function of its momentum and direction. The efficiency as a function of $s_{\mu}$
is obtained using the MC event distribution and is equal to one within 
5$\cdot$10$^{-4}$.

{\bf Tracking.}
 Using one muon to tag the presence of the other we find that the efficiency for a single muon track is about 99.6\%, resulting in a combined efficiency of 
about 99\%, almost constant in $s_\mu$. The systematic error is evaluated varying the 
purity of the control sample and ranges from 0.3 to 0.6\% as function of $s_\mu$.

{\bf Acceptance and \bbm[m_x].} 
Efficiencies for $m_x$ cuts and acceptance are
evaluated from MC, corrected to reproduce data distributions.
The systematic uncertainty due to the $m_x$ cut is obtained by moving the
cut by about one sigma  
of the mass resolution and evaluating
the difference in the $\mu\mu\gamma$ spectrum.
We find a fractional difference of 0.4\%
(constant in $s_{\mu}$) which we take as systematic error.
Systematic effects due to polar angle requirements for the
muons, $50^\circ<\theta<130^\circ$, and of dimuon, $|\cos\theta_{\mu\mu}|\!>\!\cos(15^\circ)$,
are estimated by varying the angular acceptance by $1^\circ$ 
(more than two times the resolution on the polar angle of the muon tracks)
around the nominal value. 
The systematic error ranges from 0.1 to 0.6\%.

{\bf Unfolding.} Due to the smoothness of the $\mu\mu\gamma$ spectrum and to the 
choice of bin width much larger than the mass resolution, 
resolution effects are corrected by Monte Carlo with negligible systematic error.

{\bf Software Trigger.}  A third level trigger is implemented to reduce the loss of events rejected as cosmic rays.  Its efficiency for \mmu\gam\ events, evaluated from an unbiased downscaled sample, is consistent with one within 10$^{-3}$ which is taken as systematic error.

\section{Results}\label{sec:4}
\subsection{Evaluation of $\sigma(\e^+\e^-\to \mu^+\mu^-\gamma)$  and comparison with QED at NLO}\label{subsec:4.1}
\begin{figure}[htb]\centering
  \includegraphics[height=.45\textwidth]{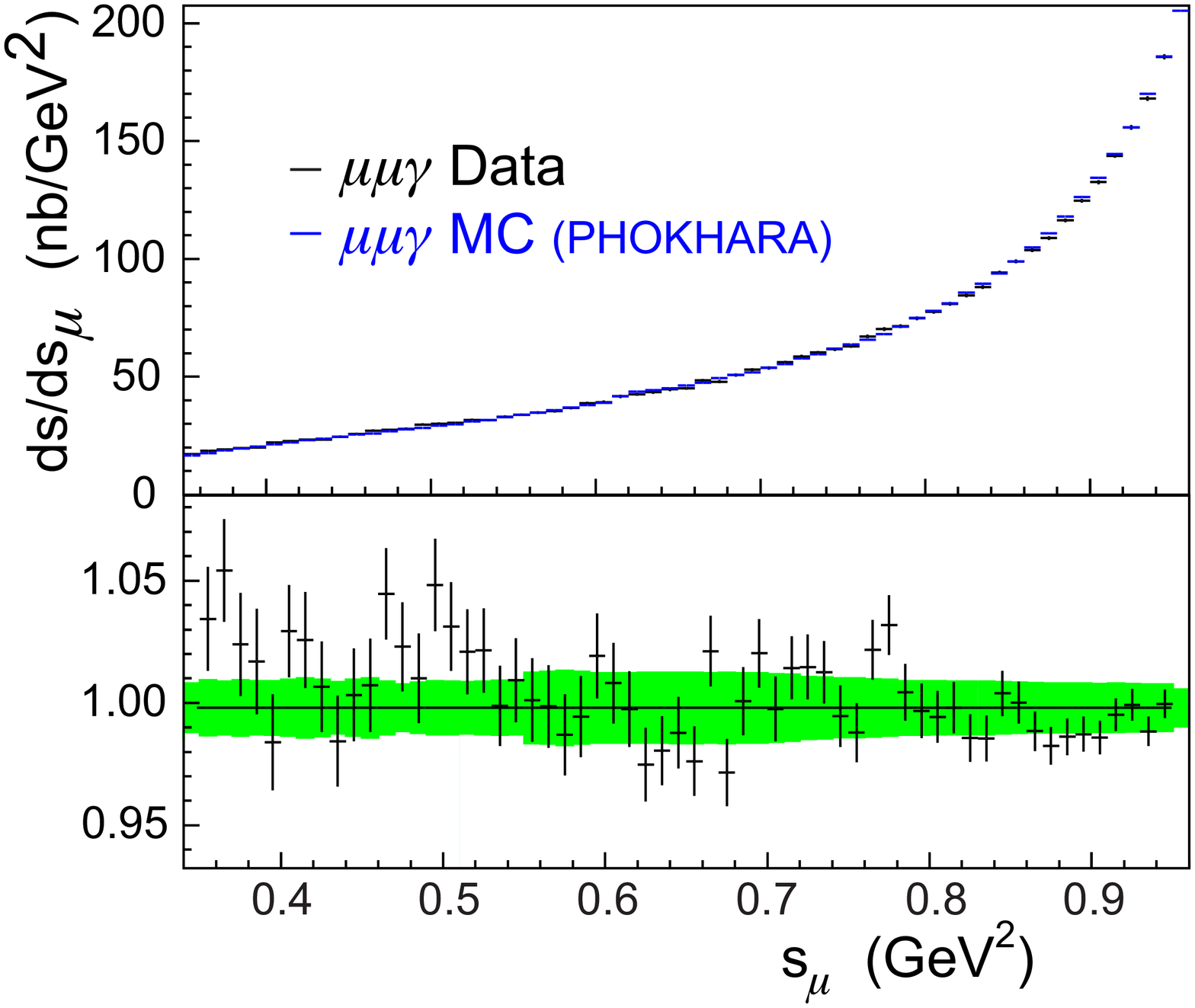}
  \caption{Top. Comparison of data and MC results for 
$\dd\sigma_{\mu\mu\gamma}/\dd s_\mu$.
   Bottom. Ratio of the two spectra. The green band shows the systematic error.}
\label{fig:4}\end{figure}

The differential $\mu^+\mu^-\gamma$ cross section is obtained from the observed event count $N_{\rm obs}$ and background  estimate $N_{\rm bkg}$,
as
\begin{equation}\frac{\dd\sigma_{\mu\mu\gamma}}
{\dd s_\mu} = \frac{N_{\rm obs}-N_{\rm bkg}}
{\Delta s_\mu}\, \frac{1}{\epsilon(s_\mu)\;\mathcal{L}}.
\label{eq:5}\end{equation}
where $\mathcal{L}$ is the integrated luminosity from Ref. \citen{Ambrosino:2006te}
and $\epsilon(s_\mu)$ the selection efficiency.
Figure \ref{fig:4} top, shows the measured $\mu^+\mu^-\gamma$ cross section
compared with the QED calculations to NLO, using the MC code
\texttt{Phokhara} \cite{Czyz:2004rj}.
Figure \ref{fig:4} bottom, shows the ratio between the two differential cross sections.
The green band indicates the systematic 
 uncertainty, experimental and theoretical, of the measured
cross section. The average ratio, using only statistical errors, is $0.9981\pm0.0015$, showing good agreement within the quoted systematic uncertainties.

\subsection{Determination of $\sigma(e^+e^-\to\pi^+\pi^-(\gam))$ from the $\pi^+\pi^-\gamma$/$\mu^+\mu^-\gamma$ ratio.}\label{subsec:4.2}
\begin{figure}[htbp]\centering
  \includegraphics[width=.7\textwidth]
  {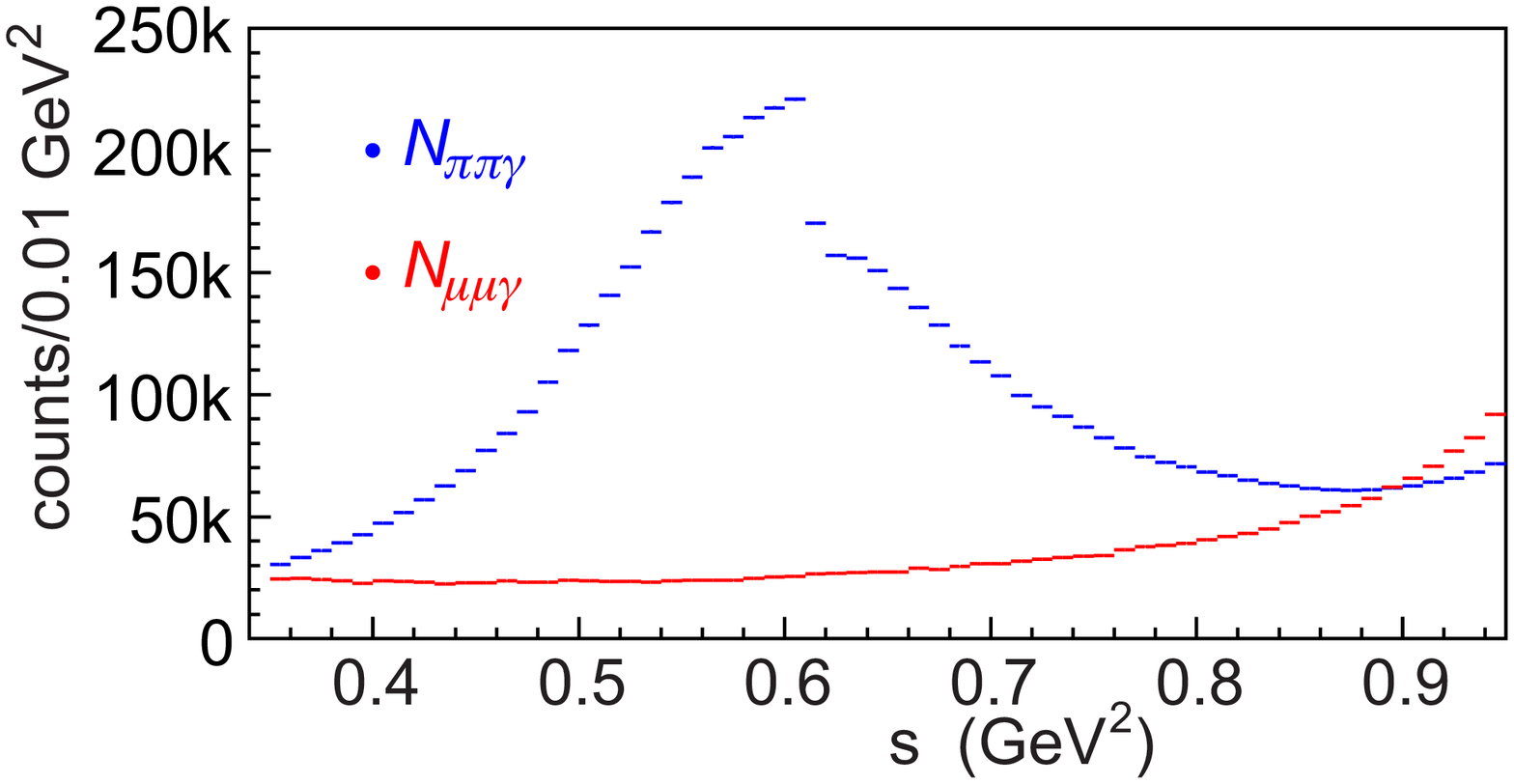}
  \caption{Square invariant mass distributions of $\pi^+\pi^-\gamma$ (blue) and
  $\mu^+\mu^-\gamma$ (red) events after background subtraction and data/MC corrections.}
\label{fig:5}\end{figure}

\begin{figure}[htbp]\centering
  \includegraphics[width=.7\textwidth]
  {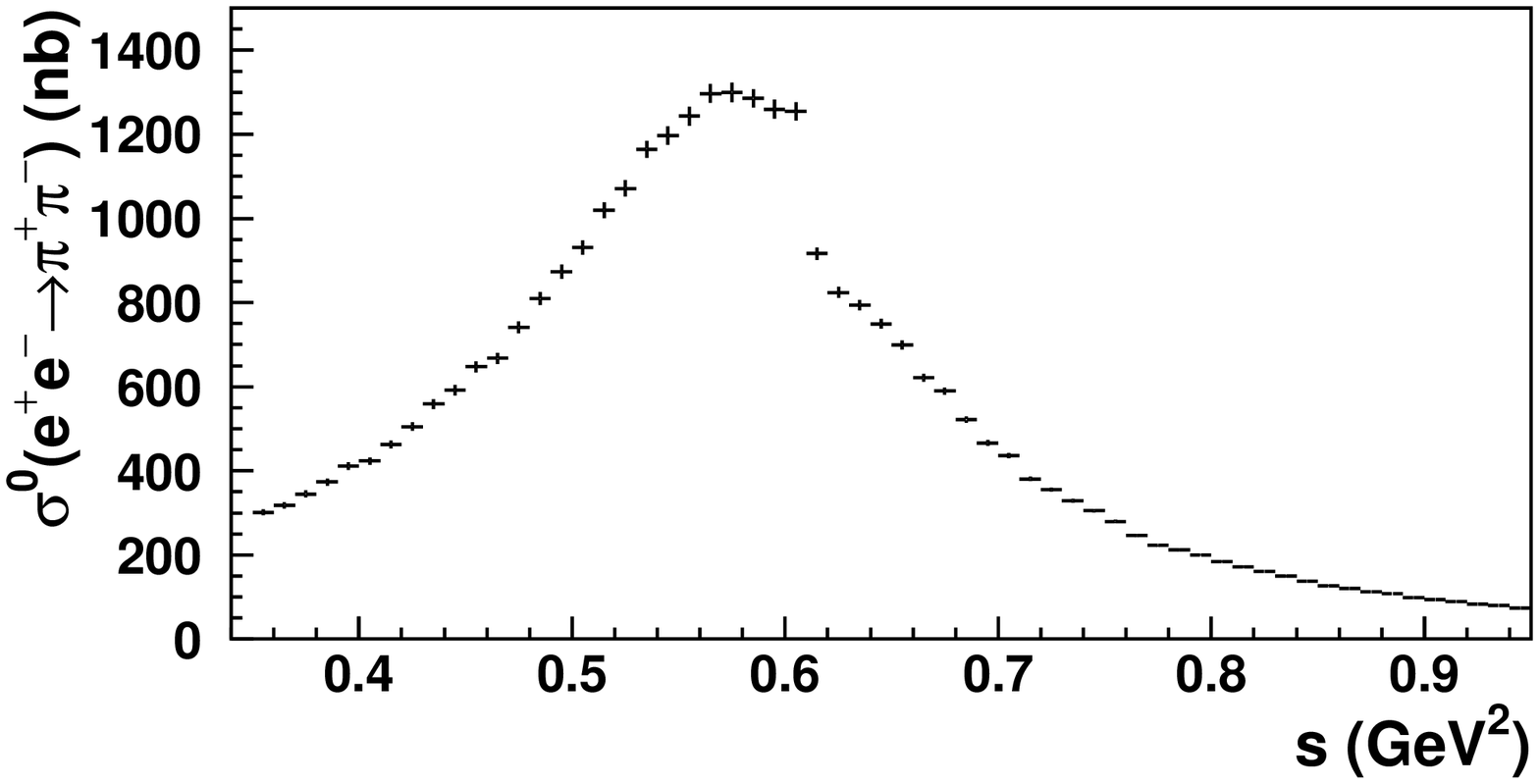}
  \caption{The bare cross section from the $\pi^+\pi^-\gamma$/$\mu^+\mu^-\gamma$ ratio.}
\label{fig:6}\end{figure}

From the bin-by-bin ratio of our published \cite{Ambrosino:2008aa} $\pi^+\pi^-\gamma$
and the $\mu^+\mu^-\gamma$ differential cross sections described above, we obtain the bare cross section $\sigma_{\pi\pi(\gamma)}^0$
(inclusive of FSR,
with VP effects removed) which is used in the dispersion integral for computing $\Delta^{\pi\pi}a_\mu$.
Figure \ref{fig:5} shows the $\pi^+\pi^-\gamma$ and $\mu^+\mu^-\gamma$ event
spectra after background subtraction and data/MC corrections.
Figure \ref{fig:6} shows the bare cross section $\sigma_{\pi\pi(\gamma)}^0$.
The pion form factor $|F_\pi|^2$ is then obtained using Eq. \ref{eq:4}.

Table~\ref{the:tab}
gives
our results for the bare cross section and the pion form factor.
Only statistical errors are shown. Systematic uncertainties on $\sigma_{\pi\pi(\gamma)}^0$
and $|F_\pi|^2$  are given in Ref. \citen{knratio}. Most of them are smaller than the 
individual 
uncertainties on $\pi\pi\gamma$ and $\mu\mu\gamma$ due to correlation between the two measurements \cite{knratio}.


\section{Evaluation of $\Delta^{\pi\pi}a_\mu$ and comparisons with other KLOE results}
\label{sec:6}
The dispersion integral for $\Delta^{\pi\pi}a_\mu$ is computed as the sum of the values for
$\sigma_{\pi\pi(\gamma)}^0$ listed in Table~\ref{the:tab} times the kernel $K(s)$, times $\Delta s=0.01$ GeV\up2 :
\begin{equation}
\Delta^{\pi\pi}a_\mu=\frac{1}{4\pi^3}\int_{s_{min}}^{s_{max}}\dd s\,\sigma_{\pi\pi(\gamma)}^0(s)\,K(s) ~,
\label{eq:7}
\end{equation}
where the kernel is given in Ref. \citen{deraf}.
Equation \ref{eq:7} gives
$\Delta^{\pi\pi}a_\mu =  (385.1\pm1.1_{\rm stat}\pm2.6_{\rm exp}\pm0.8_{\rm th})\times10^{-10}$
in the interval $0.35<M_{\pi\pi}^2< 0.95\GeV^2$.
For each 
bin contributing to the integral, statistical errors are combined in quadrature
and systematic errors are added linearly. Contributions to the $\Delta^{\pi\pi}a_\mu$ systematic uncertainty
\begin{table}[htbp]
\begin{center}
\renewcommand{\arraystretch}{1.0}
\begin{tabular}{||l|c|}
\hline\hline
Systematic sources & $\Delta^{\pi\pi}a_\mu$ \\
\hline\hline
Background subtraction & $0.6\%$ \\
Geometrical acceptance & negligible \\
$m_x$ acceptance & $0.2\%$ \\
PID & negligible \\
Tracking & $0.1\%$ \\
Trigger & $0.1\%$ \\
Unfolding & negligible \\
Software Trigger & $0.1\%$ \\
\hline
Experimental systematics & 0.7\% \\
\hline
\hline
Vacuum Polarization & negligible \\
FSR correction & $0.2\%$ \\
\hline
Theory systematics & 0.2\% \\
\hline\hline
Total systematic error & 0.7\% \\
\hline\hline
\end{tabular}
\caption{List of systematic errors on the $\Delta^{\pi\pi}a_\mu$ 
measurement.
Many systematic effects on the individual $\pi\pi\gamma$ and $\mu\mu\gamma$ analyses cancel in the ratio.}
\label{tab:4}
\end{center}
\end{table}
are shown in Table~\ref{tab:4}.
It is worth emphasizing that the use of the $\pi\pi\gamma$ to  $\mu\mu\gamma$ ratio results in a reduction of the total systematic error 
compared to the one  published in Ref. \citen{Ambrosino:2008aa} 
due to almost negligible theoretical uncertainty and
correlations between the $\pi\pi\gamma$ and $\mu\mu\gamma$ measurements~\cite{knratio}:

\vspace{-0.2cm}
{\bf Background subtraction}. The systematic uncertainty is dominated by the data/MC resolution correction in the $\mu\mu\gamma$ analysis (see Section~\ref{subsec:3.3}). Other contributions from fitting function and residual background are correlated 
for $\mu\mu\gamma$ and  $\pi\pi\gamma$  bringing the total systematic error to 
0.6\% on $\Delta^{\pi\pi}a_\mu$.\\
{\bf Geometrical and $m_x$ acceptance}. The use of the same angular cuts for the  $\pi\pi\gamma$ and $\mu\mu\gamma$ analyses yields a negligible acceptance correction in the ratio. The systematic uncertainty on  $m_x$ scale calibration is 0.2\% due to cancellations in the ratio.\\
{\bf Tracking}. The tracking-efficiency corrections are very similar for pions and muons, leading to an overall correction in the ratio which  ranges between
~0.2\% for $M^2_{\mu\mu}$ above 0.5 GeV$^2$ to 0.5\% below it. The corresponding systematic uncertainty  is conservatively estimated as 50\% of the correction value, fully bin-to-bin correlated, and translates to a 0.1\% systematic uncertainty on  $\Delta^{\pi\pi}a_\mu$.\\

The entry ``FSR correction'' in Table 2 takes into account the uncertainty on possible additional photons in the unshifting
procedure for pions~\cite{Ambrosino:2008aa} and on missing diagrams 
in \texttt{Phokhara} for $\mu\mu\gamma$. 

\begin{figure}[htb]\centering
\parbox{.5\textwidth}{ \includegraphics[width=.5\textwidth]
  {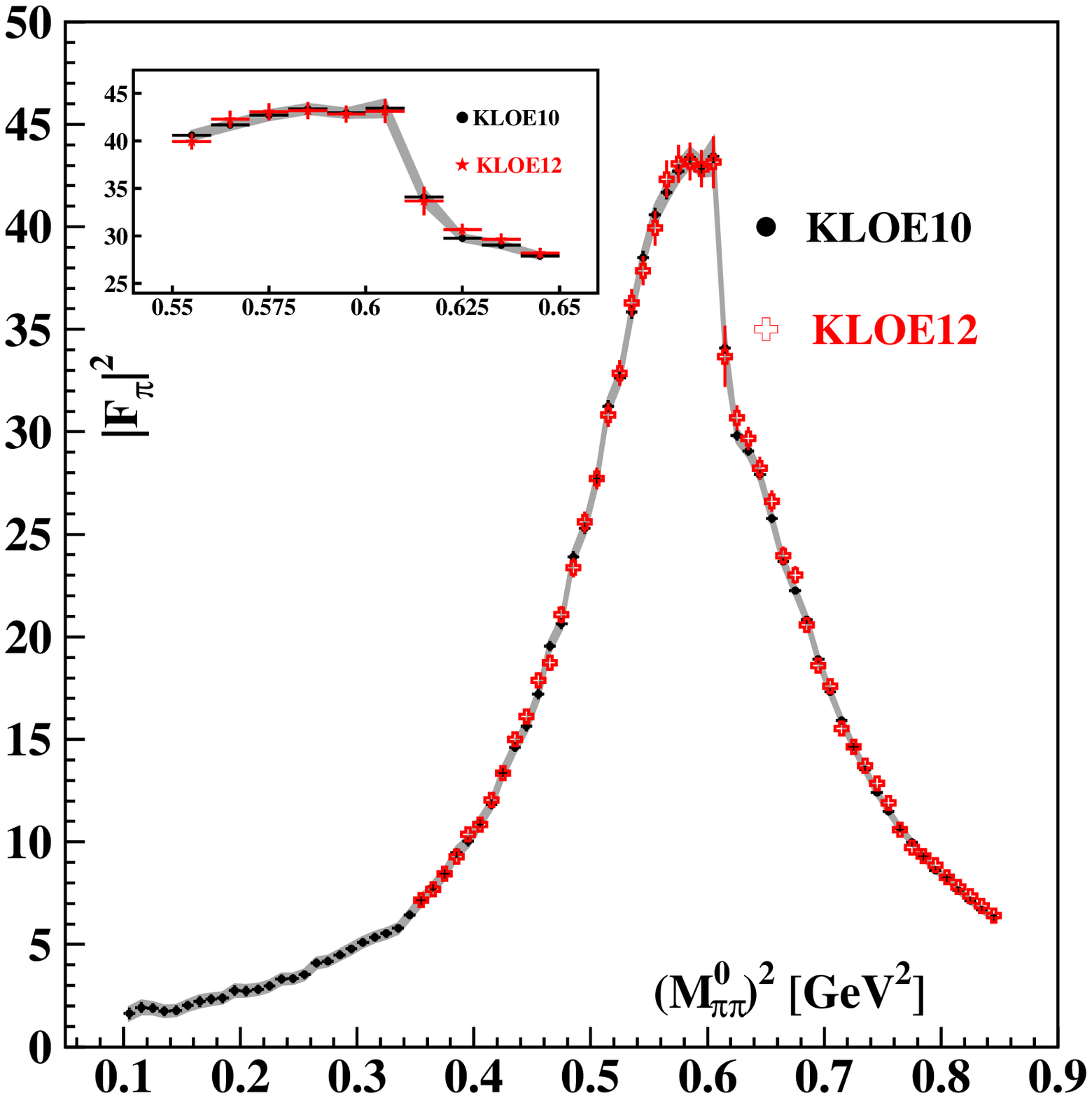}}\parbox{.5\textwidth}{ \includegraphics[width=.5\textwidth]
  {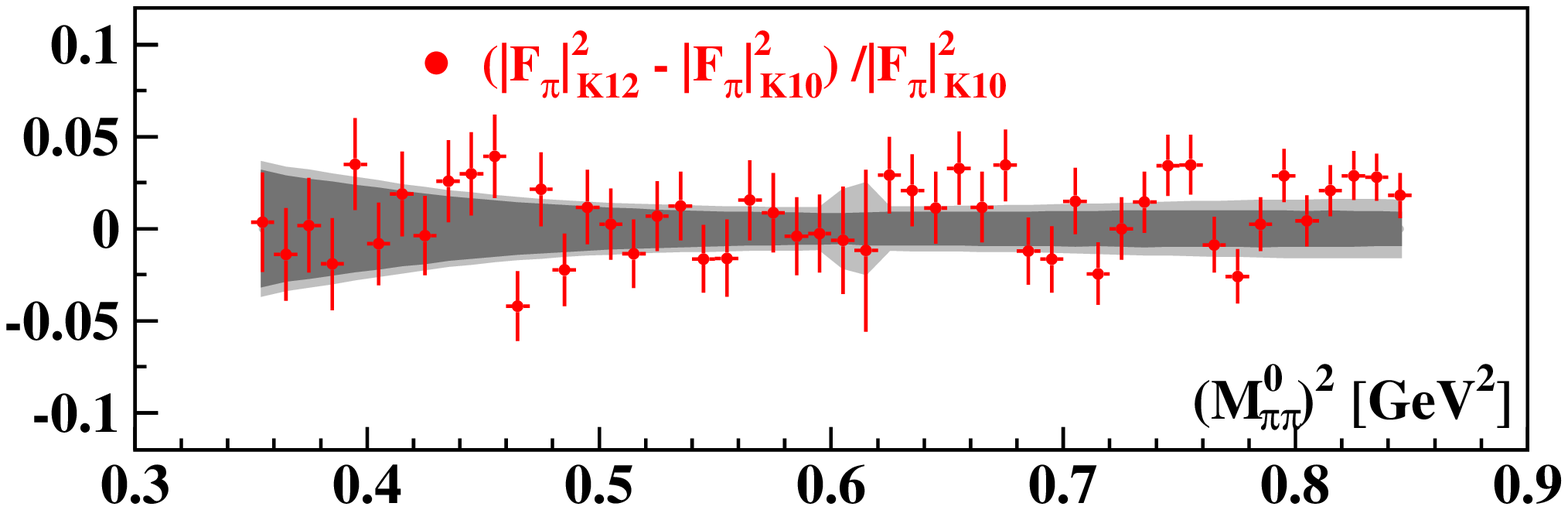}}
  \caption{
  Left: the pion form factor obtained in this work, KLOE12 (crosses) and
 from the measurement with the photon at large angle, KLOE10~\cite{Ambrosino:2010bv} (points). Right: fractional difference between the
  two $|F_\pi|^2$ measurements. The dark grey band is the statistical error from~\cite{Ambrosino:2010bv}, the light
  grey band is the combined statistical and systematic uncertainty. In both figures, errors on crosses include combined statistical and systematic uncertainties.}
\label{fig:7}\end{figure}

Figure \ref{fig:7} (top) shows the comparison between the present $|F_\pi|^2$ measurement and the previous KLOE \cite{Ambrosino:2010bv} measurement, requiring the ISR photon to be reconstructed at large angle, inside the EMC barrel. Figure \ref{fig:7} (down)
shows the fractional difference between the two measurements. They are done on
independent data sets, with different running conditions ($W= M_{\phi}$ here, $W= 1$ GeV in Ref. \citen{Ambrosino:2010bv}), and 
also with a
 different selection,
that in turn imply independent systematic uncertainties. 
The two measurements are
in very good agreement.

Table 3
 summarizes the comparison between the two most recent KLOE published measurements and
the present work on $\Delta^{\pi\pi}a_\mu$: the three results are in very good agreement.
\begin{table}[htb]
\centering\renewcommand{\arraystretch}{1.1}\begin{tabular}{r|c}
Measurement & $\Delta^{\pi\pi}a_\mu [0.35\mbox{ -- }0.85\GeV^2] \times 10^{10} $ \\\hline \hline
This work  & $377.4 \pm 1.1_{\rm stat} \pm 2.7_{\rm sys\& theo}$ \\
Large $\gamma$ angle~\cite{Ambrosino:2010bv} & $376.6 \pm 0.9_{\rm stat} \pm 3.3_{\rm sys\& theo}$ \\\hline\hline
      & $\Delta^{\pi\pi}a_\mu [0.35\mbox{ -- }0.95\GeV^2] \times 10^{10} $ \\\hline \hline
This work  & $385.1 \pm 1.1_{\rm stat} \pm 2.7_{\rm sys\& theo}$ \\
Small $\gamma$ angle~\cite{Ambrosino:2008aa} & $387.2 \pm 0.5_{\rm stat} \pm 3.3_{\rm sys\& theo}$ \\
\hline\label{tab:6}\end{tabular}
\caption{Comparison of $\Delta^{\pi\pi}a_\mu$ between our most recent measurements and the present work.}
\end{table}

\begin{table}[htbp]
\centering
\renewcommand{\arraystretch}{1.1}
\renewcommand{\tabcolsep}{5mm}
{\small
\begin{tabular}{|c|c|c||c|c|c|}
\hline
\kma$M^2_{\pi\pi}$\km&$\sigma^0_{\pi\pi(\gamma)}$&\parbox{1cm} {\vglue3mm$|F_\pi|^2$\vglue-3mm}&%
\kma$M^2_{\pi\pi}$\km&$\sigma^0_{\pi\pi(\gamma)}$&\parbox{1cm} {\vglue3mm$|F_\pi|^2$\vglue-3mm}\\
\kma GeV$^2$\kma&   nb            &    &
\kma GeV$^2$\kma&   nb            &    \\
 \hline
\km  0.355\km\kak 301.7\plm   7.7\km\kak   7.16\plm   0.18\kma &
\km  0.655\km\kak 698.8\plm  10.8\km\kak  26.62\plm   0.41\kma\\
\km  0.365\km\kak 318.6\plm   7.7\km\kak   7.68\plm   0.19\kma &
\km  0.665\km\kak 621.2\plm   9.4\km\kak  23.96\plm   0.36\kma\\
\km  0.375\km\kak 345.0\plm   8.4\km\kak   8.45\plm   0.21\kma &
\km  0.675\km\kak 589.7\plm   8.9\km\kak  23.01\plm   0.35\kma\\
\km  0.385\km\kak 373.4\plm   9.1\km\kak   9.28\plm   0.23\kma &
\km  0.685\km\kak 521.7\plm   7.7\km\kak  20.59\plm   0.31\kma\\
\km  0.395\km\kak 410.8\plm   9.4\km\kak  10.38\plm   0.24\kma &
\km  0.695\km\kak 466.1\plm   6.8\km\kak  18.60\plm   0.27\kma\\
\km  0.405\km\kak 423.1\plm   8.9\km\kak  10.85\plm   0.23\kma &
\km  0.705\km\kak 435.8\plm   6.2\km\kak  17.58\plm   0.25\kma\\
\km  0.415\km\kak 462.8\plm   9.9\km\kak  12.05\plm   0.26\kma &
\km  0.715\km\kak 380.7\plm   5.3\km\kak  15.53\plm   0.22\kma\\
\km  0.425\km\kak 504.9\plm  10.3\km\kak  13.35\plm   0.28\kma &
\km  0.725\km\kak 355.1\plm   5.0\km\kak  14.64\plm   0.21\kma\\
\km  0.435\km\kak 558.7\plm  11.6\km\kak  14.99\plm   0.31\kma &
\km  0.735\km\kak 329.2\plm   4.5\km\kak  13.72\plm   0.19\kma\\
\km  0.445\km\kak 591.4\plm  12.2\km\kak  16.11\plm   0.33\kma &
\km  0.745\km\kak 304.9\plm   4.2\km\kak  12.84\plm   0.18\kma\\
\km  0.455\km\kak 647.0\plm  13.2\km\kak  17.89\plm   0.37\kma &
\km  0.755\km\kak 279.4\plm   3.7\km\kak  11.89\plm   0.16\kma\\
\km  0.465\km\kak 667.6\plm  12.9\km\kak  18.73\plm   0.36\kma &
\km  0.765\km\kak 246.0\plm   3.2\km\kak  10.58\plm   0.14\kma\\
\km  0.475\km\kak 740.7\plm  14.1\km\kak  21.09\plm   0.40\kma &
\km  0.775\km\kak 223.7\plm   2.9\km\kak   9.72\plm   0.13\kma\\
\km  0.485\km\kak 808.9\plm  15.6\km\kak  23.37\plm   0.45\kma &
\km  0.785\km\kak 211.9\plm   2.7\km\kak   9.30\plm   0.12\kma\\
\km  0.495\km\kak 873.1\plm  16.6\km\kak  25.60\plm   0.49\kma &
\km  0.795\km\kak 200.1\plm   2.5\km\kak   8.87\plm   0.11\kma\\
\km  0.505\km\kak 931.5\plm  17.3\km\kak  27.72\plm   0.51\kma &
\km  0.805\km\kak 184.7\plm   2.2\km\kak   8.26\plm   0.10\kma\\
\km  0.515\km\kak1019.9\plm  18.3\km\kak  30.82\plm   0.55\kma &
\km  0.815\km\kak 172.6\plm   2.0\km\kak   7.80\plm   0.09\kma\\
\km  0.525\km\kak1071.4\plm  19.1\km\kak  32.87\plm   0.59\kma &
\km  0.825\km\kak 160.9\plm   1.8\km\kak   7.34\plm   0.08\kma\\
\km  0.535\km\kak1164.7\plm  20.2\km\kak  36.28\plm   0.63\kma &
\km  0.835\km\kak 149.3\plm   1.6\km\kak   6.87\plm   0.08\kma\\
\km  0.545\km\kak1196.6\plm  21.3\km\kak  37.86\plm   0.68\kma &
\km  0.845\km\kak 137.6\plm   1.5\km\kak   6.40\plm   0.07\kma\\
\km  0.555\km\kak1242.8\plm  22.1\km\kak  39.94\plm   0.71\kma &
\km  0.855\km\kak 126.5\plm   1.3\km\kak   5.94\plm   0.06\kma\\
\km  0.565\km\kak1296.8\plm  22.8\km\kak  42.33\plm   0.75\kma &
\km  0.865\km\kak 120.5\plm   1.2\km\kak   5.71\plm   0.06\kma\\
\km  0.575\km\kak1299.8\plm  22.9\km\kak  43.08\plm   0.76\kma &
\km  0.875\km\kak 112.6\plm   1.1\km\kak   5.38\plm   0.05\kma\\
\km  0.585\km\kak1285.4\plm  22.3\km\kak  43.19\plm   0.75\kma &
\km  0.885\km\kak 107.5\plm   1.0\km\kak   5.18\plm   0.05\kma\\
\km  0.595\km\kak1259.6\plm  22.3\km\kak  42.83\plm   0.76\kma &
\km  0.895\km\kak  98.6\plm   0.9\km\kak   4.80\plm   0.04\kma\\
\km  0.605\km\kak1254.3\plm  21.4\km\kak  43.15\plm   0.74\kma &
\km  0.905\km\kak  93.7\plm   0.8\km\kak   4.60\plm   0.04\kma\\
\km  0.615\km\kak 916.3\plm  14.9\km\kak  33.68\plm   0.55\kma &
\km  0.915\km\kak  88.6\plm   0.7\km\kak   4.38\plm   0.04\kma\\
\km  0.625\km\kak 824.1\plm  13.4\km\kak  30.68\plm   0.50\kma &
\km  0.925\km\kak  82.7\plm   0.7\km\kak   4.13\plm   0.03\kma\\
\km  0.635\km\kak 794.3\plm  12.1\km\kak  29.68\plm   0.45\kma &
\km  0.935\km\kak  79.6\plm   0.6\km\kak   4.00\plm   0.03\kma\\
\km  0.645\km\kak 749.0\plm  11.7\km\kak  28.23\plm   0.44\kma &
\km  0.945\km\kak  74.0\plm   0.5\km\kak   3.75\plm   0.03\kma\\
\hline
\end{tabular}}\vglue1mm
\caption{\label{the:tab}Bare cross section and the pion form factor, in $0.01\GeV^2$ intervals.
The value given in the $M_{\pi\pi}^2$ column indicates the bin center.}
\end{table}
%

\section{Conclusions}
\label{sec:7}
We have measured the differential cross section $\dd\sigma(e^+e^-\to\mu^+\mu^-\gamma)/\dd M_{\mu\mu}^2$ 
using events with
initial state radiation photons emitted at small angle and inclusive of final state radiation.
The measurement is in good agreement with QED to NLO prediction.
We determined the pion form factor from the ratio between the $\dd\sigma(e^+e^-\to\pi^+\pi^-\gamma)/\dd M_{\pi\pi}^2$
and $\dd\sigma(e^+e^-\to\mu^+\mu^-\gamma)/\dd M_{\mu\mu}^2$ cross sections, measured
with the same data set. In this way, the radiator $H$ function is not used, the luminosity of
the sample cancels out and the acceptance corrections compensate, resulting in an almost negligible systematic error.\\
This pion form factor determination is in very good agreement with previous KLOE results.
We compute the $\pi^+\pi^-$ contribution to the muon anomaly in the interval $0.592\km\,<\km\, M_{\pi\pi}\km\,<\km\,0.975$ GeV to be:
$$
\Delta^{\pi\pi}a_\mu =  (385.1\pm1.1_{\rm stat}\pm2.6_{\rm sys\,exp}\pm0.8_{\rm sys\,th})\times10^{-10} .
$$
with an experimental accuracy of 0.7\% and a theoretical uncertainty at the 0.2\% level.

This result, with comparable total experimental uncertainty and a theoretical error reduced by about 70\% with respect
to our previous results, confirms the current discrepancy between the standard model prediction (as obtained when $e^+ e^-$ data~\cite{Akhmetshin:2006bx,Akhmetshin:2006wh,Akhmetshin:2003zn,Achasov:2006vp,Aubert:2009ad,Lees:2012cj} are used) and the experimental value of  $a_\mu$.

\section*{Acknowledgments}

 We would like to thank Henryk Czy{\.z}, Sergiy Ivashyn,
Fred Jegerlehner, Johann K\"uhn, Guido Montagna, Fulvio Piccinini, Germ{\'a}n Rodrigo, Olga Shekhovtsova and Thomas Teubner for numerous useful discussions.

We thank the DA$\Phi$NE team for their efforts in maintaining low
background running conditions and their collaboration during all
data taking.
We want to thank our technical staff:
G.~F.~Fortugno and F.~Sborzacchi for their dedication in ensuring
efficient operation of the KLOE computing facilities;
M.~Anelli for his continuous attention to the gas system and detector
safety;
A.~Balla, M.~Gatta, G.~Corradi and G.~Papalino for electronics
maintenance;
M.~Santoni, G.~Paoluzzi and R.~Rosellini for general detector support;
C.~Piscitelli for his help during major maintenance periods.
This work was supported in part
by the EU Integrated Infrastructure Initiative HadronPhysics Project
under contract number RII3-CT-2004-506078;
by the European Commission under the 7th Framework Programme through
the 'Research Infrastructures' action of the 'Capacities' Programme,
Call: FP7-INFRASTRUCTURES-2008-1, Grant Agreement N. 227431;
by the Polish National Science Centre through the Grants No.
0469/B/H03/2009/37, 0309/B/H03/2011/40, 2011/01/D/ST2/00748, 
2011/03/N/ST2/02641, 2011/03/N/ST2/02652   and by the
Foundation for Polish Science through the MPD programme and the
project HOMING PLUS BIS/2011-4/3.

\end{document}